\documentclass{iaus}
\usepackage{graphicx}

\title[short title of paper] 
{Velocity Fields of Spiral Galaxies in z$\sim$0.5 Clusters}

\author[Kutdemir E. et al.]   
{Elif Kutdemir$^{1,2}$, %
Bodo Ziegler$^1$ \break \and Reynier F. Peletier$^2$}

\affiliation{$^1$Georg-August-Universit\"{a}t, Institut f\"{u}r Astrophysik, Friedrich-Hund-Platz 1, 37077
G\"{o}ttingen, Germany \break email: kutdemir@astro.physik.uni-goettingen.de\\[\affilskip]
$^2$Kapteyn Astronomical Institute, P.O. BOX 800, 9700AV, Groningen, 
The Netherlands}

\pubyear{2007}
\volume{241}  
\pagerange{1--2}
\date{?? and in revised form ??}
\jname{Proceedings Title IAU Symposium}
\editors{R.F. Peletier, A. Vazdekis, eds.}
\begin{document}

\maketitle

\begin{abstract} Spiral galaxies can be affected by interactions in clusters, that
also may distort the internal velocity field. If unrecognized from single-slit 
spectroscopy, this could lead to a wrong determination of the maximum  rotation
velocity as pointed out by \cite[Ziegler \etal\ 2003]{Bodo03}. This parameter 
directly enters into the Tully--Fisher relation, an important tool to  investigate
the evolution of spiral galaxies. To overcome this problem, we  measure the
2D-velocity fields by observing three different slit positions  per galaxy using
FORS2 at the VLT providing us with full coverage of each galaxy and an adequate
spatial resolution. The kinematic properties are compared to structural features
determined on the HST/ACS images to assess possible interaction processes. As a
next step, the whole analysis will be performed for three more clusters, so that 
we will be able to establish a high-accuracy TFR for spirals at z$\sim$0.5.
\end{abstract}


{\bf Overview: }We describe here the analysis of both gaseous and stellar
kinematics of $\sim$20 spiral galaxies in the cluster MS0451-03 at $z=0.54$
explaining the steps that have been done so far to get a Tully--Fisher
relation for that cluster.

\begin{figure}
 \resizebox{13.5cm}{!}{\includegraphics{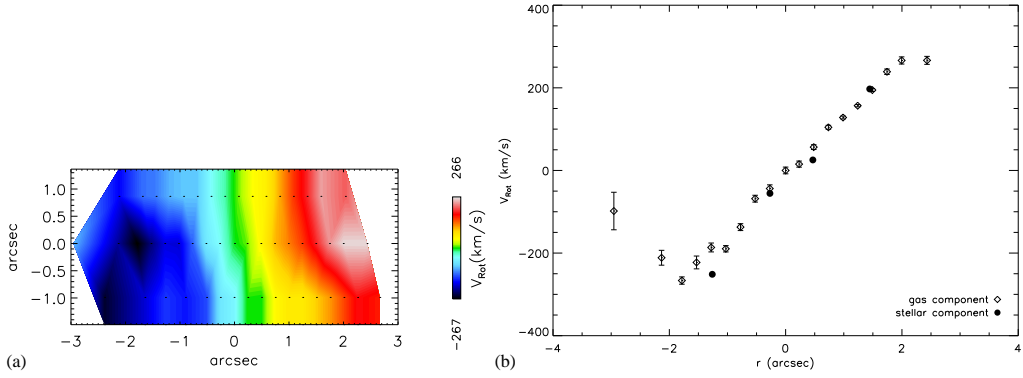}}
 \centering
  \caption{(\textit{a}) The [OII] velocity map of a galaxy in cluster MS0451-03
  constructed using the position-velocity diagrams obtained from 3 slits
  (\textit{b}) Stellar and [OII] gas position-velocity diagrams of the same
  galaxy plotted on top of each other}\label{vfield}
\end{figure}

{\bf Gas Velocity Fields: } Spectra of three slits parallel to the photometric
axis of each galaxy were obtained using MXU masks with FORS2 at the VLT
exhibiting several emission  lines. For each line and for each slit position, a
position-velocity diagram was extracted. This information was converted to a
single coordinate system to construct the velocity field taking into account
the small difference in the positioning of the 3 masks compared to each other
(figure~\ref{vfield}\,(\textit{a})).

\begin{figure}
 \resizebox{13.5cm}{!}{\includegraphics{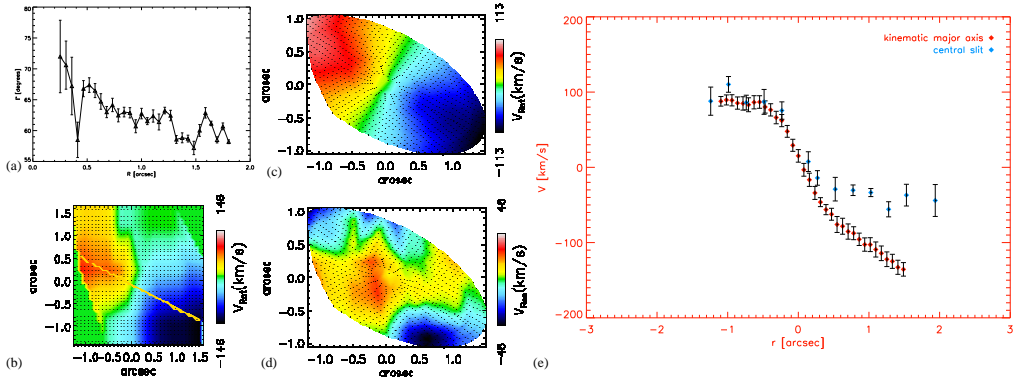}}
  \caption{Some results from a kinemetric analysis of the H$\beta$ velocity field of a background galaxy at z$\approx$0.58. 
  (\textit{a}) The kinematic position angle as a function of radius
  (\textit{b}) The kinematic major axis (defined as the median PA) plotted on the velocity map
  (\textit{c}) A simple 2 dimensional kinematic fit (obtained fixing PA \& ellipticity and excluding the higher order Fourier terms)
  (\textit{d}) Difference between the observed velocity map and the simple kinematic fit
  (\textit{e}) The position-velocity diagram extracted along the central slit
  together with the position-velocity diagram extracted along the kinematic major axis}\label{kin}
\end{figure}

{\bf Kinematic Analysis: }First, we determined the central coordinates, using the assumption that
the velocity gradient is maximum at the kinematic center. The kinemetry method (\cite[Krajnovi\'{c}
\etal\ 2006]{Kra06}) then allows to derive the kinematic axis, that can be different from the
photometric one (figure~\ref{kin}). We quantify deviations from circular motion in inclined disks
using a high-order Fourier analysis.

{\bf Photometric Analysis: }Structural parameters were derived on HST/ACS images
in the $I$ band fitting the 2D surface brightness of each galaxy using GALFIT
(\cite[Peng \etal\ 2002]{Peng02}). These are needed for the derivation of the
maximum rotation velocities. For each galaxy we fitted a S\'{e}rsic bulge and
an exponential disk. Residual images are analyzed to look for spiral arms, bars
and tidal features. Exploiting additional ground based photometry, we determine
the rest frame $B$ band luminosity, the second parameter of the Tully--Fisher
relation.

{\bf Analyzing Stellar Component: }For some of the galaxies we were able to
measure the strength and position of strong absorption lines. For these, we have also
extracted the stellar rotation curves (figure~\ref{vfield}\,(\textit{b})). This
was done using the PPXF software (\cite[Capellari \& Emsellem 2004]{ce04}) and the
MILES stellar library (\cite[S\'{a}nchez-Bl\'{a}zquez \etal\ 2006]{sa06}). From
the best fitting mix of stellar templates, we can also investigate the star
formation histories (average ages) of the stellar populations of those galaxies.

{\bf Acknowledgements: }We thank the DFG \& VolkswagenStiftung for financial support and Da Rocha, 
Kronberger \& Verdugo for discussion.

\end{document}